\documentstyle[eqsecnum,aps,prd,preprint,psfig]{revtex}
\begin{document}

\draft

\title{Inflation and Nonsingular Spacetimes of Cosmic Strings}
\author{Inyong Cho\footnote[1]{Electronic address: cho@cosmos2.phy.tufts.edu}}

\address{Institute of Cosmology,
        Department of Physics and Astronomy,\\
        Tufts University,
        Medford, Massachusetts 02155, USA}
\date{\today}
\maketitle

\begin{abstract}
Inflation of cosmic gauge and global strings is investigated
by numerically solving the combined Einstein and field equations.
Above some critical symmetry-breaking scales, 
the strings undergo inflation along the radial direction
as well as the axial direction at the core.
The nonsingular nature of the spacetimes around supercritical gauge 
and global strings is discussed and contrasted to the singular
static solutions that have been discussed in the literature.
\end{abstract}


\section{Introduction}

Cosmic strings are linelike topological defects 
that may form as a result of a phase 
transition in the early universe.
If a string is associated with a magnetic field, it is called a gauge string,
otherwise it is a global string.
They have attracted much attention because of their cosmological importance:
deficit angle in the spacetime geometry and a candidate for the seed of
structure formation in the early universe.

It was proposed that topological defects can inflate if the 
symmetry-breaking scale satisfies 
$\eta \gtrsim \eta_c \sim {\cal O}(m_p)$ in Refs.~\cite{Linde,Vilenkin}. 
This was later verified in numerical simulations by 
Sakai {\it et al.}~\cite{Sakai}.
They found, in particular, that the critical value of $\eta$ for domain walls
and global monopoles is $\eta_c \simeq 0.33m_p$.
Then what about cosmic strings? There is no reason why we exclude cosmic 
strings out of the topological inflationary category.
Recently, it was numerically proved that a (2+1) dimensional gauge string 
can inflate by de Laix {\it et al.}~\cite{Tanmay}.

In this paper, we shall numerically solve the combined Einstein and
field equations for a gauge and a global string in
(3+1) spacetime dimensions.
For the gauge string, we find that the core inflates if $\eta \gtrsim 0.25m_p$
with unit winding number in the critical coupling case (Bogomol'nyi limit).
For the global string, $\eta \gtrsim 0.23m_p$.
The critical values decrease as the winding number increases. For the 
gauge string, the critical value also decreases slightly as the coupling 
of the gauge field to the scalar field becomes weaker than the self
coupling of the scalar field.

The asymptotic spacetime of a gauge string is known to be 
conical~\cite{VilenkinC}.
This spacetime exhibits a deficit angle $\Delta =8\pi G\mu$, 
where $\mu\sim \eta^2$ is the mass per unit length of the string.
When the symmetry-breaking scale is sufficiently large,
the deficit angle exceeds $2\pi$ and analyses of the static 
solution show that the spacetime possesses a physical singularity
outside the core of the string~\cite{Gott,Ortiz,Laguna}.
However, from our numerical results we know that 
supermassive strings are dynamical and undergo inflation at the core.
Therefore, we believe that the static treatment of supermassive strings
loses its validity and that we should treat them in a time-dependent way.

For global strings, the singularity exists regardless of 
the symmetry-breaking scale.
Many people have tried to find a static solution of a global string and 
they found that there also exists a physical singularity outside
the core of the string~\cite{Cohen,Sikivie,Gregorys,Gibbons}.
What was suggested to remove this singularity is again a time-dependent
treatment of the string. 
Gregory~\cite{Gregoryt} introduced a specific metric which has 
an axial time dependence  and showed that this spacetime is nonsingular.

In our work, we follow the evolution of 
supermassive gauge and global strings in a general
time-dependent metric and show that no singularity 
develops in the spacetimes around the strings.

In the next section, we solve the Abelian Higgs model of a gauge string
and discuss its inflation and spacetime geometry. 
Sec.~III is devoted to global strings.
Our conclusions are summarized in Sec.~IV.
In Appendix, we show the equations in detail and the numerical algorithms.

\section{Gauge string}

Let us consider the Abelian Higgs model of a string with a complex scalar
field $\Phi$ and a $U(1)$ gauge field $A_\mu$, coupled to gravity.
The action is

\begin{equation}
S = \int d^4x \sqrt{-g} \left( {{\cal R} \over 16\pi G} + {\cal L}
\right)\,,
\label{eq=action}
\end{equation}
where ${\cal R}$ is Ricci scalar and the Lagrangian ${\cal L}$ is

\begin{equation}
{\cal L} = -D_\mu \Phi {\bar D}^\mu {\bar \Phi}
	-{1 \over 4}F_{\mu\nu}F^{\mu\nu} -V(\Phi {\bar\Phi})\,.
\label{eq=lagr}
\end{equation}
Here the covariant derivative is $D_\mu = \nabla_\mu +ieA_\mu$, 
the gauge field strength is $F_{\mu\nu} = \nabla_\mu A_\nu - \nabla_\nu A_\mu$,
and the potential is
\[
V(\Phi {\bar\Phi}) = {\lambda \over 4}(\Phi {\bar\Phi}-\eta^2)^2\,,
\]
where $e$ is the coupling constant between gauge and scalar fields,
$\lambda$ is the self-coupling constant of the scalar field, and 
$\eta$ is the symmetry-breaking scale of the scalar field.
We use a general cylindrically symmetric metric

\begin{equation}
ds^2 = -dt^2 + B(t,r)^2 dr^2 +C(t,r)^2r^2d\theta^2 + H(t,r)^2 dz^2\,.
\label{eq=metric}
\end{equation}
(We use $\hbar = c = 1$, $G=1/m_p^2$.)
In these coordinates, we can use a time-dependent generalization
of the usual Nielsen-Olesen anzatz for the scalar and gauge fields

\[
\Phi = \phi(t,r) e^{in\theta}\,,
\]
\[
A_\mu = -{n \over e}\alpha(t,r)\nabla_\mu \theta\,,
\]
where $n$ is the winding number of the string.
We also set the usual boundary conditions

\begin{eqnarray} 
\phi(t,0)=0\,,\qquad \phi(t,\infty)=\eta\,,\nonumber\\
\alpha(t,0)=0\,,\qquad \alpha(t,\infty)=1\,.
\label{eq=bc}
\end{eqnarray}
For the above action, the Einstein's equation is
\[
G_{\mu\nu} = 8\pi GT_{\mu\nu}\,,
\]
where the energy-momentum tensor is given by

\begin{equation}
T_{\mu\nu} = D_\mu \Phi {\bar D}_\nu {\bar \Phi}
	+D_\nu \Phi {\bar D}_\mu {\bar \Phi} 
	+g^{\alpha\beta}F_{\mu\alpha}F_{\nu\beta}
	+g_{\mu\nu}{\cal L}\,.
\label{eq=emtensor}
\end{equation}
The scalar field equation is

\[
D_\mu D^\mu \Phi = {\partial V(\Phi {\bar\Phi}) \over \partial {\bar\Phi}}\,,
\]
and the gauge field equation is

\[
\nabla^\nu F_{\mu\nu} = ie( {\bar\Phi}\nabla_\mu \Phi
	-\Phi\nabla_\mu {\bar\Phi}) -2e^2A_\mu\Phi {\bar\Phi}\,.
\]
The detailed equations are given in the Appendix.

The static solution at a large distance from the core is

\begin{equation}
ds^2 = -dT^2 +dR^2 +(1-4G\mu)^2R^2d\theta^2 +dZ^2\,,
\label{eq=static}
\end{equation}
where $\mu$ is the mass per unit length of the string and proportional
to $\eta^2$. The metric (\ref{eq=static}) exhibits a deficit angle
$\Delta = 8\pi G\mu$.
As $\eta$ increases, the deficit angle also increases. 
When the deficit angle exceeds $2\pi$, the static solution (\ref{eq=static})
ceases to exist and we expect that the string becomes dynamical and undergoes
inflation at the core in this regime.

Let us consider the role of the winding number for inflation.
The coupling between scalar and gauge fields is given by

\[
\beta \equiv \left( {m_s \over m_v} \right)^2 = {\lambda \over 2e^2} \,,
\]
where $m_s$ and $m_v$ are masses of scalar and gauge fields.
In the limit of critical coupling (Bogomol'nyi limit), $\beta=1$,
the mass per unit length is $\mu =2\pi |n|\eta^2$~\cite{Linet}.
Then the critical value of $\eta$ at which the static solution 
(\ref{eq=static}) ceases to exist is proportional to $1/\sqrt{n}$.
Therefore, as the winding number increases, the core of the string can inflate
at lower symmetry-breaking scales.
Another way to explain the inflation of topological defects is the core
size being bigger than the horizon scale~\cite{Linde,Vilenkin}.
Under this condition the gravitational effect becomes important
in the core region and the defects are supposed to inflate.
The energy density of the string can be roughly estimated by

\[
\rho \sim H^2 \sim {\mu \over \delta^2} \sim {n\eta^2 \over \delta^2}\,,
\]
where $\delta$ is the core size and $H$ is the Hubble parameter.
The core size is then

\begin{equation}
{\delta \over H^{-1}} \sim \sqrt{n}\eta\,.
\label{eq=coresize}
\end{equation}
The core of the string can inflate when its size becomes 
comparable to the horizon scale, $\delta/H^{-1} \sim 1$.
This condition gives again $\eta_c \sim 1/\sqrt{n}$ from 
Eq.~(\ref{eq=coresize}).
In other words, for a given symmetry-breaking scale $\eta$, 
the core size $\delta$ is proportional to $\sqrt{n}$.
Therefore, as the winding number increases, the wider core of the string
can inflate at lower symmetry-breaking scales.
Figure~\ref{fig=ggnsize} shows the core sizes as functions of the
winding number for several $\eta$'s in the flat spacetime.

One important effect of the gauge field coupled to the scalar field 
is that it makes the string well localized and, thereby, suppresses the 
energy divergence along the radial direction. 
As the coupling constant $\beta$
increases, the effect of the gauge field becomes weaker, therefore,
the core size becomes bigger. 
The string has a better chance to inflate 
with the bigger core acquired as $\beta$ increases.

We numerically solve the field equations introduced earlier.
Initially we assume a flat spacetime and obtain 
the initial scalar and gauge fields,
$\phi(0,r)$ and $\alpha(0,r)$, by numerically solving the flat-spacetime
field equations. 
We then solve the time-dependent field equations
to trace the evolution of these fields as well as the gravitational
field $g_{\mu\nu}$.
We define the core size by the proper radius $Cr$ at which the scalar
field is $\phi=\eta/2$ and determine whether inflation occurs
by observing the growth of this proper radius.
For $n=1$ and $\beta=1$, the critical value of $\eta$ for inflation 
is found to be $\eta_c \simeq 0.25m_p$. 
This value is remarkably close to the critical value $0.255m_p$ found
by Laguna and Garfinkle~\cite{Laguna}, 
at which the static solution becomes singular.
The critical value decreases as $n$ and $\beta$ increase as expected.
The critical values of $\eta$ are given in Table~\ref{tab=ggeta}
for several $n$'s and $\beta$'s.
Figure~\ref{fig=ggsfd} shows the evolution of the scalar field configuration
as a function of the proper radius. 
$t$ and $r$ are scaled in the unit of
the horizon size at the center, $H_0^{-1}=[8\pi GV(\phi=0)/3]^{-1/2}$.
We can see clearly that the proper radius grows very rapidly (exponentially)
in the core region.

Having found the inflationary behavior in the supermassive string\footnote{
``Supermassive'' was named for the gauge string to which the static
solution (\ref{eq=static}) no longer applies},
it is worth while to discuss its spacetime.
For the supermassive string, the static solution (\ref{eq=static}) is 
no longer valid.
However, since the energy-momentum tensor of the string falls off
rapidly outside the core, the asymptotic form of the metric must be
one of the two Levi-Civita metrics~\cite{Civita},

\begin{equation}
ds^2 = -dt^2 +dz^2 +dr^2 +(a_1r+a_2)^2d\theta^2
\label{eq=levi}
\end{equation}
which is conical, or

\begin{equation}
ds^2 = (b_1R+b_2)^{4/3}(-dT^2 + dZ^2)
	+dR^2 + (b_1R + b_2)^{-2/3}d\theta^2\,,
\label{eq=kasner}
\end{equation}
which is a special case of a Kasner metric~\cite{Kasner}.
The qualities of the metrics (\ref{eq=levi}) and (\ref{eq=kasner}) 
depend on the signs of the constants $a_1$, $a_2$, $b_1$, and $b_2$.
When $a_1$ and $a_2$ have the same sign, the metric (\ref{eq=levi})
is equivalent to the metric (\ref{eq=static}). 
For $b_1$ and $b_2$ with different signs,
Laguna and Garfinkle~\cite{Laguna} examined that the asymptotic spacetime
of the supermassive string is described by the metric (\ref{eq=kasner}).
For $a_1$ and $a_2$ with different signs,
Ortiz~\cite{Ortiz} demonstrated that the metric (\ref{eq=levi}) can also
be a solution.
In both cases, the spacetimes possess physical singularities at
$r=-a_2/a_1$ and $R=-b_2/b_1$. The singularity approaches the core
as $\eta$ increases. 
Nevertheless, it is not clear whether
it is physically sensible to have a singularity in the string spacetime.
Since we have shown that the supermassive string undergoes inflation, 
we believe that the static treatment is no longer valid and the string
should be treated in a time-dependent way to describe its
dynamical nature.
We expect that the time-dependent treatment removes the singularity
from the spacetime of the supermassive string.

One simple way to examine the existence of a physical singularity is
to analyze the metric. The Riemann tensor contains
the second derivatives of the metric terms.
If the metric terms do not contain cusps and are smooth functions
of the coordinates up to the second derivatives, 
then the scalar invariants calculated by the Riemann tensor 
are finite and it is safe to say that  no
singularity develops in the spacetime.
While we perform the numerical calculation, we follow the evolution of
the metric.
Figure~\ref{fig=ggbch} shows logarithmic values of the metric terms 
$B$, $C$, and $H$ at several moments of time. 
They are smooth functions of $r$ and change also 
smoothly in $t$.
We also calculate the Kretschmann scalar
$R_{\alpha\beta\gamma\delta}R^{\alpha\beta\gamma\delta}$.
This invariant is finite everywhere and every moment in time as
shown in Fig.~\ref{fig=ggrrcv}.
Therefore, we conclude that the time-dependent treatment 
of supermassive strings makes a singularity-free spacetime possible.
This treament is also useful to deal with the singularity of the global string
in the next section.

As inflation proceeds at the core of the string, the metric terms 
$B$, $C$, and $H$
grow rapidly and the de Sitter expansion is established around the center
of the string
\[
{{\dot B} \over B} \approx {{\dot C} \over C} \approx 
{{\dot H} \over H} \approx \sqrt{{8\pi G \over 3}V(\phi=0)}\,.
\]
In addition to radial inflation, the string also inflates
along the axial direction at the core. 
As shown in Fig.~\ref{fig=ggk31}, at the core, the ratio
${\dot{H} \over H}/{\dot{B} \over B}$ remains close to a constant 
($\approx 1$). This indicates that the expanding behavior along
the $z-$ direction is similar to that along the $r-$ direction.

\section{Global String}

In this section, we consider the model of a $U(1)$ global string.
The action, field equations, and the metric are the same as those
of the gauge string in the previous section without the gauge field
($A_\mu = 0$).
In the absence of the gauge field  the scalar field configuration
of a global string stretches further radially.
The situation is equivalent to that of the gauge string 
with $\beta\to\infty$ $(e\to 0)$.
On the basis of the cause of inflation discussed in the previous section, 
this guarantees that the global string will inflate 
at a lower symmetry-breaking scale and that
the inflationary picture should not be much different.
For $n=1$, the critical value of $\eta$ for inflation is found to be
$\eta_c\approx 0.23m_p$ which is slightly lower than that of the gauge string.
The critical value decreases as the winding number increases.
The critical values are given for several winding numbers in 
Table~\ref{tab=gbeta}.

As mentioned in the Introduction, a static global string has a physical
singularity outside the core.
In this region, the energy-momentum tensor is given by

\begin{equation}
T_0^0 =T_1^1 =T_3^3 =-T_2^2 ={\eta^2 \over g_{22}}\,,
\label{eq=gbstatem}
\end{equation}
and the closed-form solution of the Einstein equations was found by Cohen and
Kaplan~\cite{Cohen}

\begin{equation}
ds^2 =\left( {u \over u_0} \right)(-dt^2+dz^2)
	+\gamma^2 \left( {u_0 \over u} \right)^{1/2}
	e^{{u_0^2-u^2 \over u_0}}(du^2+d\theta^2)\,,
\label{eq=Cohen}
\end{equation}
where $u_0=1/8\pi G\eta^2$, and $\gamma$ is a constant of integration.
This metric is transformed to the linearized solution obtained
by Harari and Sikivie~\cite{Sikivie} when $\eta \ll m_p$,

\begin{equation}
ds^2 =\left[ 1-4G\mu(r) \right](-dt^2+dz^2) + dr^2
       +\left[ 1-8G\mu(r) \right] r^2d\theta^2\,.
\label{eq=Sikivie}
\end{equation}
Here $\mu(r)\simeq \int_{\delta}^r T_0^0 2\pi rdr \simeq 2\pi\eta^2
\ln\left( r/\delta \right)$ is the string mass per unit length
out to a distance scale $r$ from the core ($\approx\delta$).
The relation between $u$ and $r$ is 
$u \approx u_0 -\ln\left(r/\delta\right)$.
Then the boundary of the core in the metric (\ref{eq=Cohen}) 
is located at $u\simeq u_0$
and $u =\infty$ corresponds to the center of the string.
The solution (\ref{eq=Cohen}) exhibits coordinate singularities
at $u=0$ and $u=\infty$.
The one at $u=0$ turns out to be a physical one and the existence
of this singularity was also examined by analyzing the field equations
in Refs.~\cite{Gregorys,Gibbons}.
For Grand-Unified-Theory scale strings 
($\eta \sim 10^{16}\text{GeV}\sim 10^{-3}m_p$), 
the singularity is expelled outside the horizon, therefore, 
it does not cause any cosmological trouble.
However, for supermassive strings, it is located near the core and could
thus be problematic.

In a similar way to the supermassive gauge string,
one way to escape from this trouble is to introduce a time-dependent
string. Indeed, Gregory~\cite{Gregoryt} removed the singularity 
by taking a specific time-dependent form of the metric: 
only axial time dependence was introduced.
But this metric still has a coordinate singularity (not a physical one).
It was later questioned by Wang and Nogales~\cite{Wang} that
this coordinate singularity is unstable to both test particles
and physical perturbations. 
In particular, they argued that the back reaction of the perturbations
of null dust fluids will turn the coordinate singularity into
a physical one.
However, the situation is quite different for the supermassive
strings. 
Because of the dynamical nature of the supermassive strings, 
we need to solve the system with a general time-dependent metric and
scalar field.

We now solve the equations numerically as we did for the gauge string.
Fig.~\ref{fig=gbbch} shows the logarithmic values of 
the metric terms $B$, $C$, and $H$
and Fig.~\ref{fig=gbrrcv} shows the Kretschmann scalar
$R_{\alpha\beta\gamma\delta}R^{\alpha\beta\gamma\delta}$
at several moments of time. Their regular behaviors allow us to conclude
that no singularity develops in the spacetime of a global
string.
For large $\eta$ global strings like supermassive gauge strings,
the appropriate way to deal with
their dynamical nature is to use a time-dependent treatment.

\section{Conclusions}

We have investigated inflation in cosmic strings.
In the core region, the strings undergo inflation radially as well as axially
when $\eta \gtrsim \eta_c$.
With unit winding number ($n=1$) the critical values for inflation were found 
to be $\eta_c \approx 0.25m_p$ for a gauge string in  the Bogomol'nyi limit
($\beta=1$) and $\eta_c =0.23m_p$ for a global string.
The critical values decrease as $n$ and $\beta$ increase.
We have explained this $\eta_c$ variation in terms of the core size of 
defects. The core of defects inflates when its size becomes bigger than 
the horizon scale: 
for larger $n$ and $\beta$, strings have bigger cores,
and the global string has a bigger core than the gauge string
for a given $\eta$.
Regardless of the symmetry-breaking scale $\eta$, around 
the center of defects the de Sitter expansion 
is established since the scalar field
stays about the top of the potenital ($\phi\approx 0$).
However, this is not sufficient for the cores of defects to inflate. 
Inflation requires another condition which is the core size being comparable
to the horizon scale so that the core can be dynamical due to
the gravitational effect.
Or equivalently, the potential $V(\phi)$ needs to be flat enough at
$\phi\approx 0$ so that the field $\phi$ can spend enough time about the top
of the potential.
For this condition to be satisfied, the symmetry-breaking scale $\eta$
needs to be sufficiently large.
This description also explains why we have somewhat lower critical values
of $\eta$ for strings than those for domain walls and global monopoles 
($\eta_c\approx 0.33m_p$).
Strings have bigger cores at the same symmetry-breaking scale
than the other defects.

For supermassive gauge strings and all scale global strings, 
we have had troublesome physical singularities outside the core when we
treat them in a static way.
The elegant exit to nonsingular spacetimes is to introduce a time-dependent
treatment.
From the numerical simulations we could show 
that there is no singularity
developing around time-dependent supermassive strings.

\acknowledgements
Author is grateful to Alex Vilenkin, Pablo Laguna, Tanmay Vachaspati, 
and Xavier Siemens
for helpful discussions and to National Science Foundation for partial
support.

\appendix
\section*{Field equations and numerical algorithms}

Einstein's equations with the metric (\ref{eq=metric}) and 
the energy-momentum tensor (\ref{eq=emtensor}) are

\begin{eqnarray}
-G_0^0 &=& K_1^1K_2^2 +K_2^2K_3^3 +K_3^3K_1^1 \nonumber\\
       &+& {1 \over B^2}\left( -{C'' \over C} -{H'' \over H}
	+{B' \over B}{C' \over C} -{C' \over C}{H' \over H}
	+{H' \over H}{B' \over B}+{B' \over Br}
	-2{C' \over Cr} -{H' \over Hr} \right) \nonumber\\
       &=& 8\pi G \left[ \dot{\phi}^2 + {\phi'^2 \over B^2}
	+{n^2 \over C^2r^2}\phi^2 (1-\alpha)^2 
	+{n^2 \over 2e^2}{1 \over C^2r^2}\left( \dot{\alpha}^2
	+{\alpha'^2 \over B^2} \right) +V(\phi) \right]\,,
\label{eq=G00}
\end{eqnarray}
\begin{eqnarray}
{1 \over 2}(-G_1^1 +G_2^2 +G_3^3) &=& \dot{K_1^1} -(K_1^1)^2
	-{1 \over 2}(K_1^1K_2^2 -K_2^2K_3^3 +K_3^3K_1^1) \nonumber\\
	&+& {1 \over 2B^2}\left( {C'' \over C} +{H'' \over H}
	-{B' \over B}{C' \over C} -{C' \over C}{H' \over H}
	-{H' \over H}{B' \over B} -{B' \over Br}
	+2{C' \over Cr} -{H' \over Hr} \right) \nonumber\\
	&=& 4\pi G \left[ \dot{\phi}^2 -3 {\phi'^2 \over B^2}
	+{n^2 \over C^2r^2}\phi^2 (1-\alpha)^2 
	-{n^2 \over 2e^2}{1 \over C^2r^2}\left( \dot{\alpha}^2
	+{\alpha'^2 \over B^2} \right) -V(\phi) \right]\,,
\label{eq=K11}
\end{eqnarray}  	
\begin{eqnarray}
{1 \over 2}(G_1^1 -G_2^2 +G_3^3) &=& \dot{K_2^2} -(K_2^2)^2
	-{1 \over 2}(K_1^1K_2^2 +K_2^2K_3^3 -K_3^3K_1^1) \nonumber\\
	&+& {1 \over 2B^2}\left( {C'' \over C} -{H'' \over H}
	-{B' \over B}{C' \over C} +{C' \over C}{H' \over H}
	+{H' \over H}{B' \over B} -{B' \over Br}
	+2{C' \over Cr} +{H' \over Hr} \right) \nonumber\\
	&=& 4\pi G \left[ \dot{\phi}^2 + {\phi'^2 \over B^2}
	-3{n^2 \over C^2r^2}\phi^2 (1-\alpha)^2 
	+{n^2 \over 2e^2}{1 \over C^2r^2}\left( 3\dot{\alpha}^2
	-{\alpha'^2 \over B^2} \right) -V(\phi) \right]\,,
\label{eq=K22}
\end{eqnarray}  	
\begin{eqnarray}
{1 \over 2}(G_1^1 +G_2^2 -G_3^3) &=& \dot{K_3^3} -(K_3^3)^2
	-{1 \over 2}(-K_1^1K_2^2 +K_2^2K_3^3 +K_3^3K_1^1) \nonumber\\
	&+& {1 \over 2B^2}\left( -{C'' \over C} +{H'' \over H}
	+{B' \over B}{C' \over C} +{C' \over C}{H' \over H}
	-{H' \over H}{B' \over B} +{B' \over Br}
	-2{C' \over Cr} +{H' \over Hr} \right) \nonumber\\
	&=& 4\pi G \left[ \dot{\phi}^2 + {\phi'^2 \over B^2}
	+{n^2 \over C^2r^2}\phi^2 (1-\alpha)^2 
	-{n^2 \over 2e^2}{1 \over C^2r^2}\left( \dot{\alpha}^2
	-3{\alpha'^2 \over B^2} \right) -V(\phi) \right]\,,
\label{eq=K33}
\end{eqnarray}
\begin{eqnarray}
G_{01} &=& K_2^{2\prime} +K_3^{3\prime}
	-(K_1^1 -K_2^2)\left( {C' \over C} +{1 \over r} \right)
	-(K_1^1 -K_3^3){H' \over H} \nonumber\\
	&=& 2\dot{\phi}\phi' +{n^2 \over e^2}{1 \over C^2r^2}
	\dot{\alpha}\alpha'\,, 
\label{eq=G01}
\end{eqnarray}
where
\begin{equation}
K_1^1 = -{\dot{B} \over B}\,,\qquad K_2^2 = -{\dot{C} \over C}\,,\qquad
K_3^3 = -{\dot{H} \over H}\,.
\label{eq=Ks}   	
\end{equation}
The field equation for $\phi$ is
\begin{equation}
\ddot{\phi} -(K_1^1 +K_2^2 +K_3^3)\dot{\phi} -{\phi'' \over B^2}
+{1 \over B^2}\left( {B' \over B} -{C' \over C} -{H' \over H}
-{1 \over r} \right)\phi' +{n^2 \over C^2r^2}\phi (1-\alpha)^2
+{\lambda \over 2}\phi (\phi^2 -\eta^2) =0\,.
\label{eq=sfd}
\end{equation}
The field equation for $\alpha$ is
\begin{equation}
\ddot{\alpha} -(K_1^1 -K_2^2 +K_3^3)\dot{\alpha} -{\alpha'' \over B^2}
+{1 \over B^2}\left( {B' \over B} +{C' \over C} -{H' \over H} +{1 \over r}
\right)\alpha' +2e^2\phi^2(\alpha -1) =0\,.
\label{eq=vfd}
\end{equation}
The field equations for the global string are obtained by setting 
$\alpha =0$.
At $t=0$, we assume a flat spacetime, $B(0,r)=C(0,r)=H(0,r)=1$, 
and zero velocities of
scalar and gauge fields, $\dot{\phi}(0,r) = \dot{\alpha}(0,r) =0$.
Then we solve the scalar and vector field equations (\ref{eq=sfd}) and
(\ref{eq=vfd}) to obtain $\phi(0,r)$ and $\alpha(0,r)$.
$K_1^1(0,r)$ and $K_2^2(0,r)$ are evaluated by the Hamiltonian and
momentum constraint equations (\ref{eq=G00}) and (\ref{eq=G01}) 
after setting $K_3^3(0,r)=0$.
In the next time step, $B(t,r)$, $C(t,r)$, and $H(t,r)$ are calculated 
by Eq.~(\ref{eq=Ks}).
The $K_i^i(t,r)$'s are calculated by Eqs.~(\ref{eq=K11})-(\ref{eq=K33}), 
and $\phi(t,r)$ and $\alpha(t,r)$ are calculated by Eqs.~(\ref{eq=sfd})
and (\ref{eq=vfd}).
The boundary conditions for $\phi$ and $\alpha$ are given in Eq.~(\ref{eq=bc}).
We also use the boundary condition $H'(t,0)=0$ ($K_3^{3\prime}=0$) and
apply regularity condition $K_1^1(t,0) = K_2^2(t,0)$ at the origin.
To solve the field equations numerically, 
we use the modified Crank-Nicholson method.
We insist that the Hamiltonian and momentum constraint equations,
(\ref{eq=G00}) and (\ref{eq=G01}), are satisfied at every moment although
they are not solved directly.

We have only two constraint equations to set initial three $K_i^i(0,r)$'s.
So we assume $K_3^3(0,r)=0$ which is consistent with the Einstein's equations.
Nonetheless, it might not be the best choice. However, even if this choice
is not the optimal one to describe our physical system, the numerical system
will find its correct route propagating with the speed of light
from the center of the string.
The region in which we are interested 
relaxes to the correct configuration
during the period of our numerical iteration.

\begin{table}
\begin{tabular}{ccccc}
n & 1 & 2 & 3 & 4 \\ \hline
$\eta_c /m_p$ & $0.25\pm 0.005$ & $0.17\pm 0.005$ &
$0.13\pm 0.005$ & $0.11\pm 0.005$
\end{tabular}
\begin{center}
(a)
\end{center}
\begin{tabular}{cccccc}
$\beta$ & 0.5 &1 & 2 & 3 & 4 \\ \hline
$\eta_c /m_p$ & $0.255\pm 0.0025$ & $0.25\pm 0.0025$ & $0.24\pm 0.0025$ &
$0.235\pm 0.0025$ & $0.235\pm 0.0025$
\end{tabular}
\begin{center}
(b)
\end{center}
\caption{
(a) The critical values of $\eta$ for the gauge string
in the Bogomol'nyi limit $\beta=1$.
(b) The critical values of $\eta$ for $n=1$.}
\label{tab=ggeta}
\end{table}

\vspace{1.in}

\begin{table}
\begin{tabular}{cccccc}
n & 1 & 2 & 3 & 4 & 5 \\ \hline
$\eta_c /m_p$ & $0.23\pm 0.0025$ & $0.155\pm 0.0025$ &
$0.115\pm 0.0025$ & $0.095\pm 0.0025$ & $0.075\pm 0.0025$
\end{tabular}
\vspace{.5in}
\caption{
The critical values of $\eta$ for the global string}
\label{tab=gbeta}
\end{table}

\begin{figure}
\caption{
A plot of $\delta/H_0^{-1}$ vs. $n$
for the $\eta=0.1,0.5,1.0,2.0,3.0m_p$ gauge strings
(from the bottom to the top) in the flat spacetime.
The dashed line corresponds to $\sqrt{n}$ as a reference.
$H_0^{-1}$ is the horizon size at the center of the string.
}
\label{fig=ggnsize}
\end{figure}

\begin{figure}
\caption{
The scalar field configurations as functions of the proper radius
$CH_0r$ at $H_0t=0,2,4$ (from the left to the right)
for the $\eta=0.5m_p$ gauge string ($n=1$, $\beta=1$).
The rapid growth of the proper radius in the core region
indicates inflation.
}
\label{fig=ggsfd}
\end{figure}

\begin{figure}
\caption{
Plots of (a) $\log_{10}B$, $\log_{10}C$, and (b) $\log_{10}H$ vs. $H_0r$
at $H_0t=1,3,5$ (from the bottom to the top) for the $\eta=0.5m_p$
gauge string ($n=1$, $\beta=1$).
The metric terms behave regularly in $r$ and $t$.
}
\label{fig=ggbch}
\end{figure}

\begin{figure}
\caption{
A plot of $R_{\alpha\beta\gamma\delta}R^{\alpha\beta\gamma\delta}/H_0^4$
vs. $H_0r$ at $H_0t=2,4,6$ (from the right to the left)
for the $\eta=0.5m_p$ gauge string ($n=1$, $\beta=1$).
The scalar invariant is finite everywhere.
The generic picture is not very different for different $\eta$'s.
}
\label{fig=ggrrcv}
\end{figure}

\begin{figure}
\caption{
A plot of ${\dot{H} \over H}/{\dot{B} \over B}$ vs. $H_0r$ 
at $H_0t=2,3,4,5$ (from the bottom to the top)
for the $\eta=0.5m_p$ gauge string ($n=1$, $\beta=1$).
The ratio stays close to a constant ($\approx 1$) in the core region 
($H_0r\protect\lesssim 0.5$).
}
\label{fig=ggk31}
\end{figure}

\begin{figure}
\caption{
Plots of (a) $\log_{10}B$, $\log_{10}C$, and (b) $\log_{10}H$ vs. $H_0r$
at $H_0t=1,3,5$ (from the bottom to the top) for the $\eta=0.4m_p$
global string ($n=1$).
The metric terms behave regularly in $r$ and $t$.
}
\label{fig=gbbch}
\end{figure}

\begin{figure}
\caption{
A plot of $R_{\alpha\beta\gamma\delta}R^{\alpha\beta\gamma\delta}/H_0^4$
vs. $H_0r$ at $H_0t=2,4,6$ (from the right to the left)
for the $\eta=0.4m_p$ global string ($n=1$).
The scalar invariant is finite everywhere.
The generic picture is not very different for different $\eta$'s.
The core is located at $H_0r \protect\lesssim 0.5$.
}
\label{fig=gbrrcv}
\end{figure}

\newpage
\begin{figure}
\psfig{file=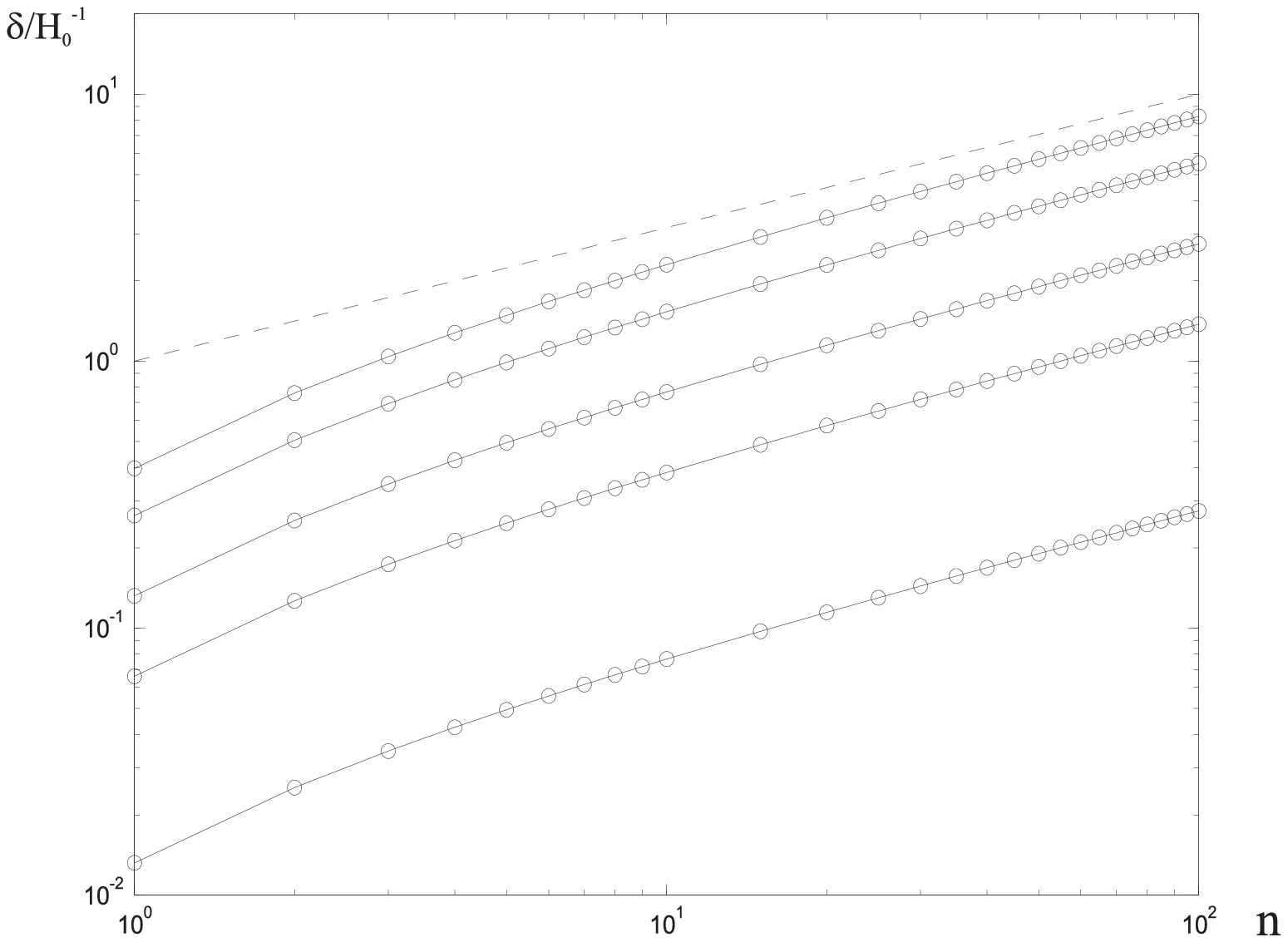}
\end{figure}

\newpage
\begin{figure}
\psfig{file=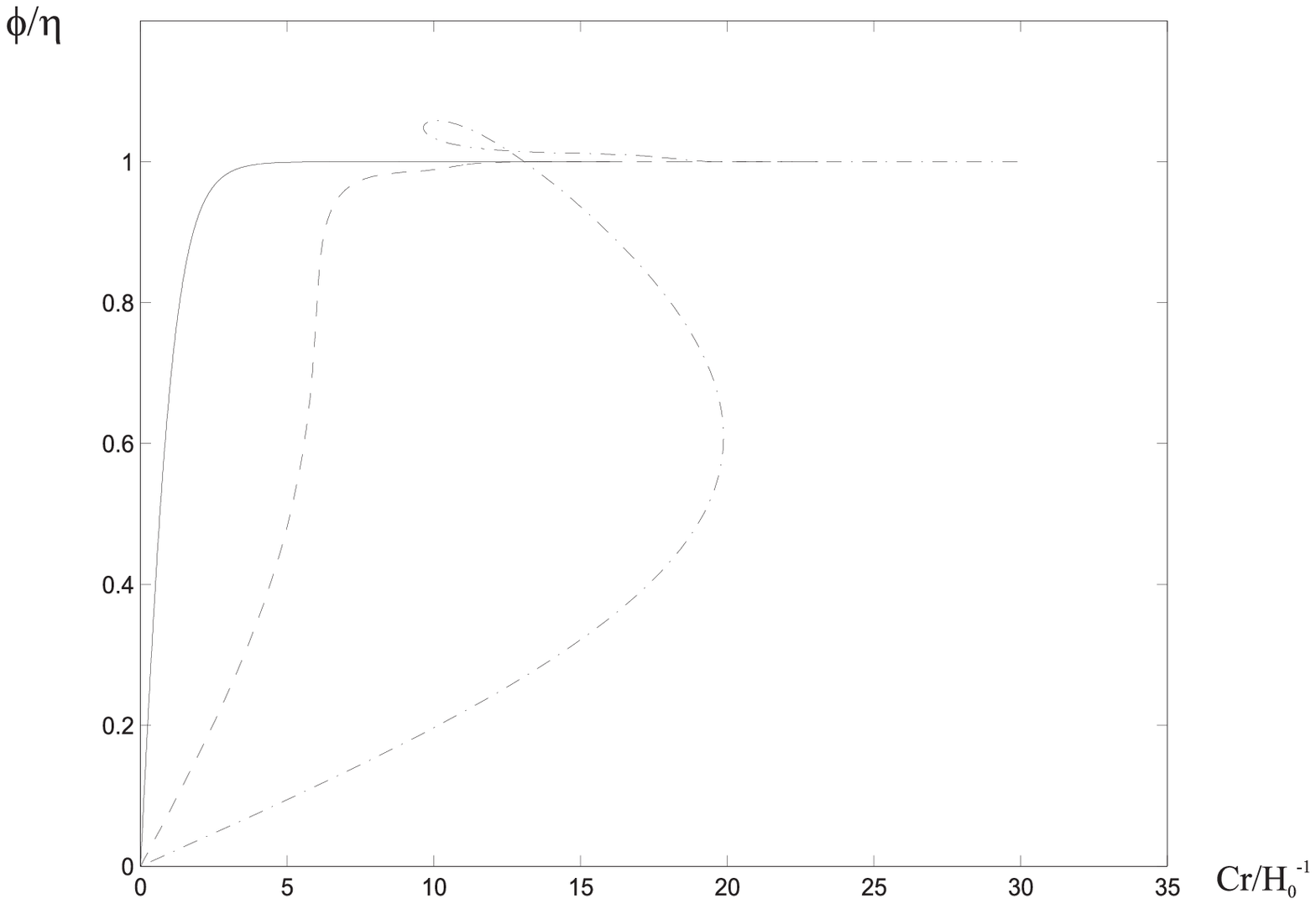}
\end{figure}

\newpage
\begin{figure}
\psfig{file=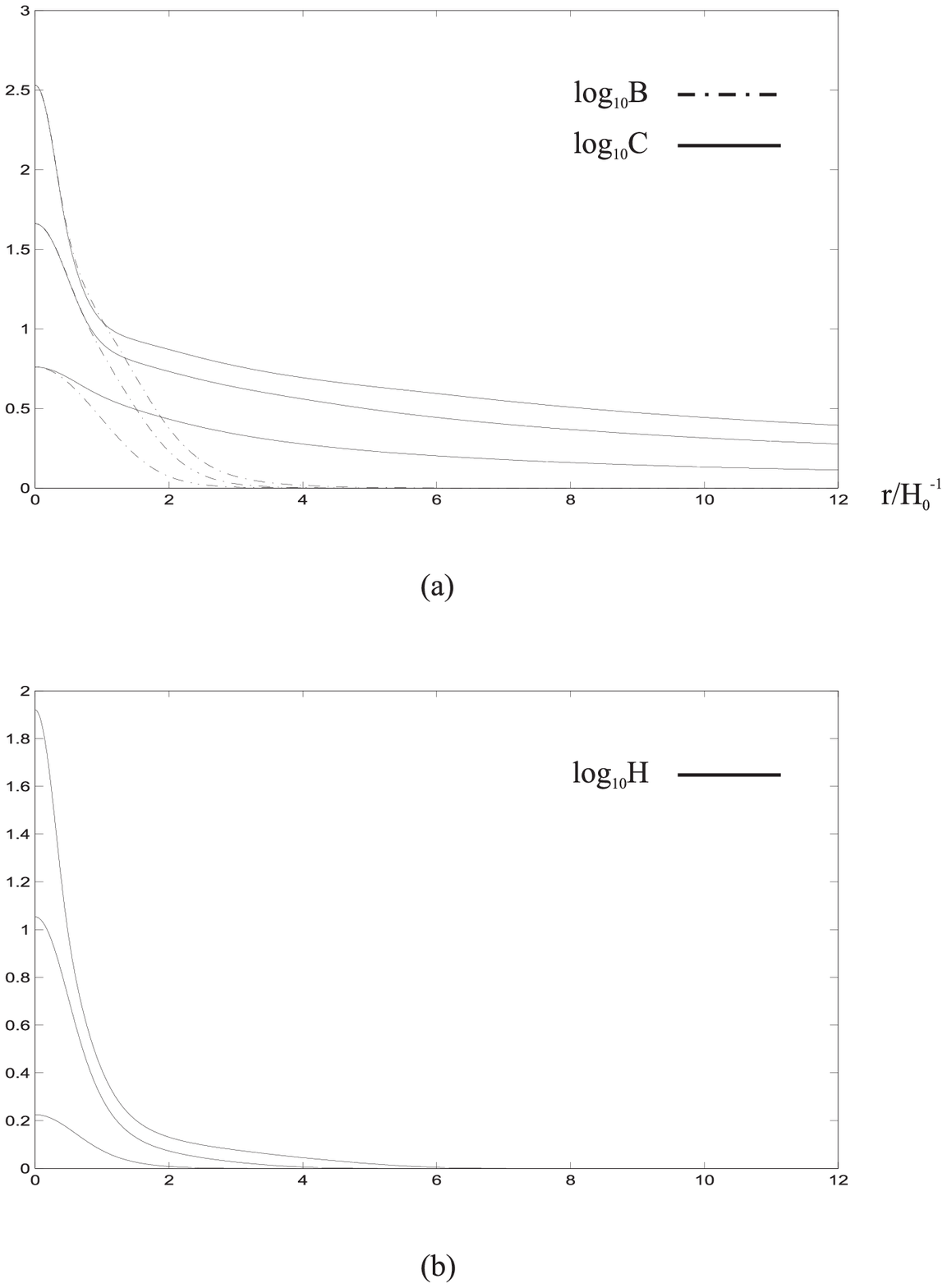}
\end{figure}

\newpage
\begin{figure}
\psfig{file=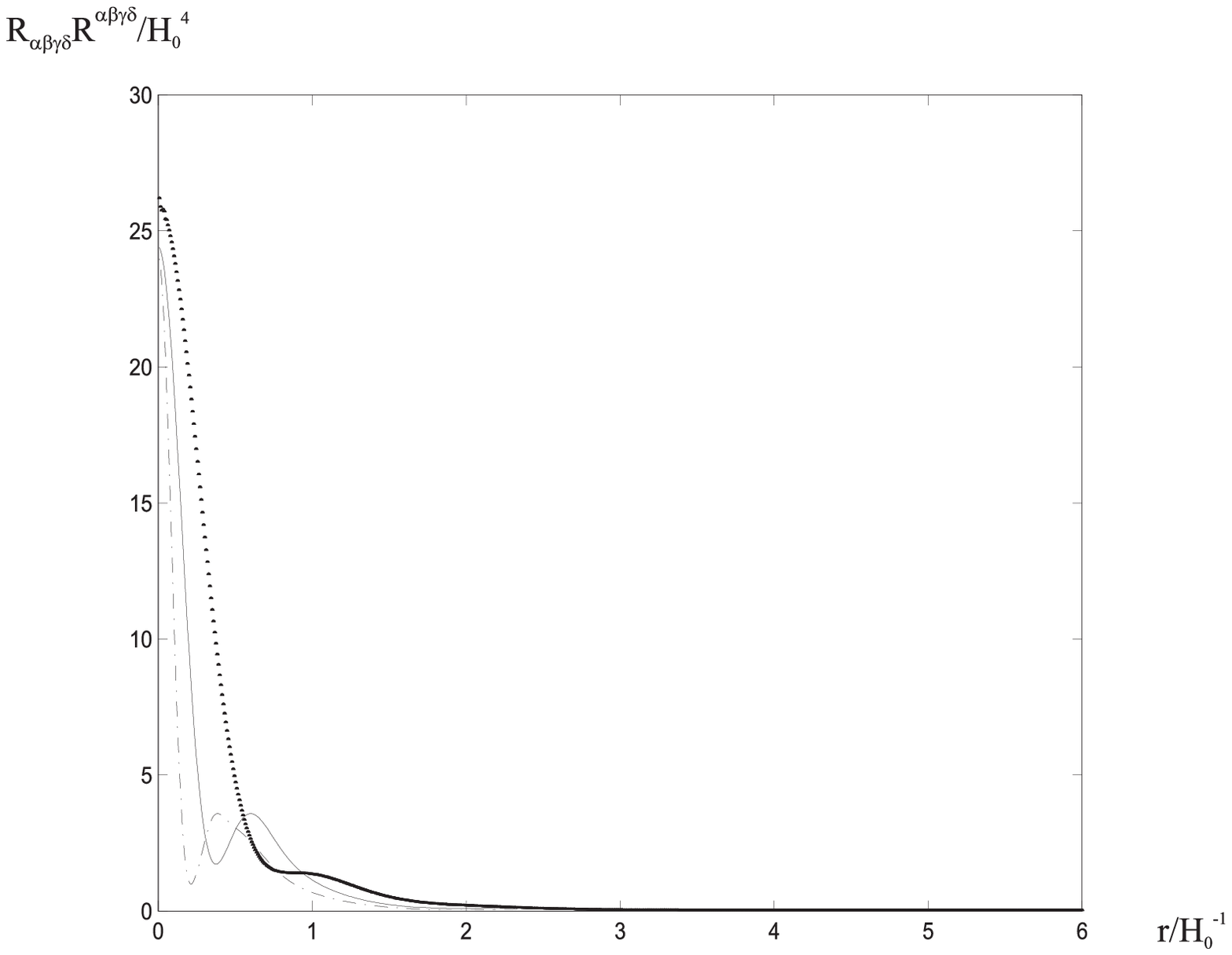}
\end{figure}

\newpage
\begin{figure}
\psfig{file=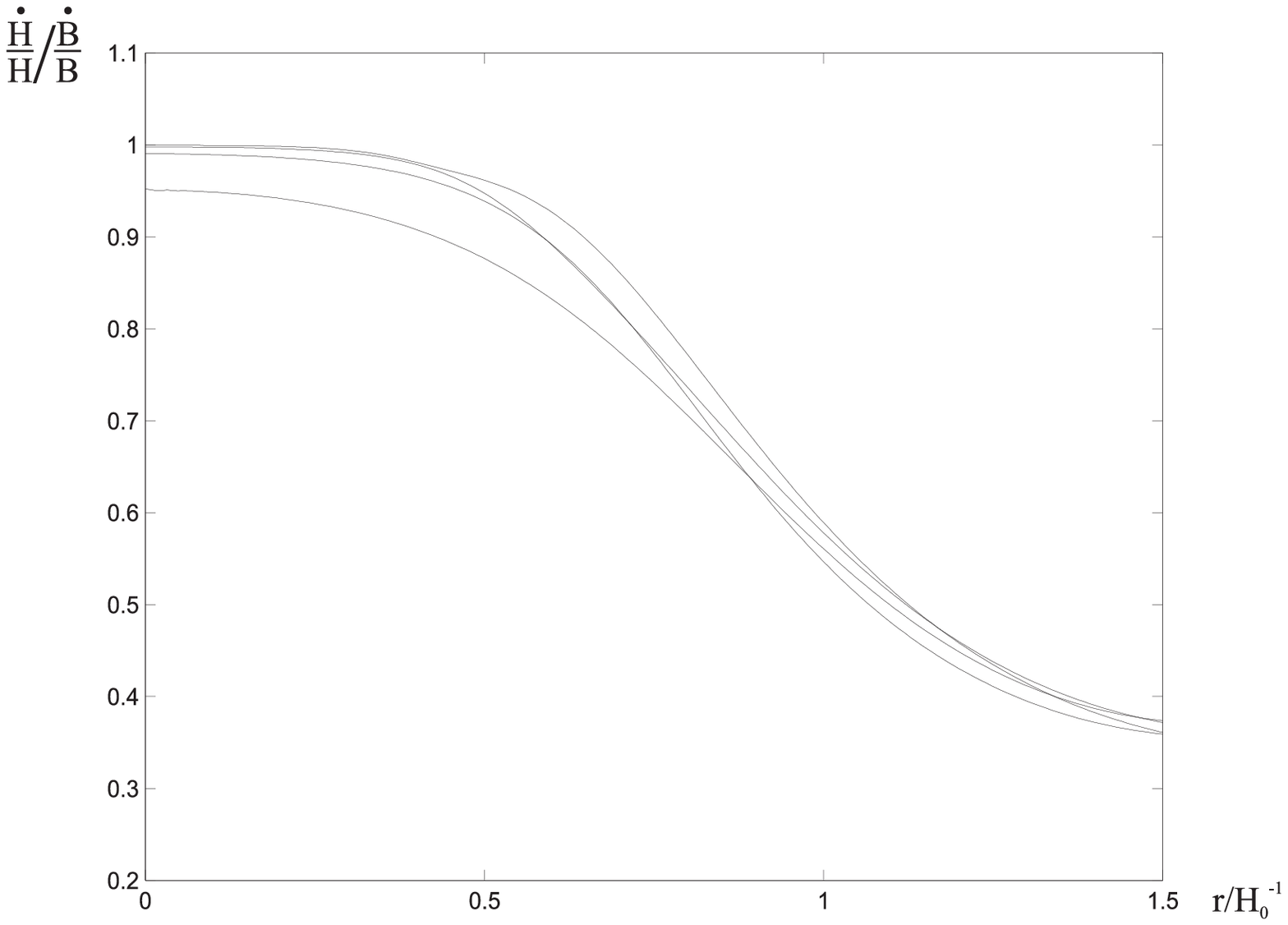}
\end{figure}

\newpage
\begin{figure}
\psfig{file=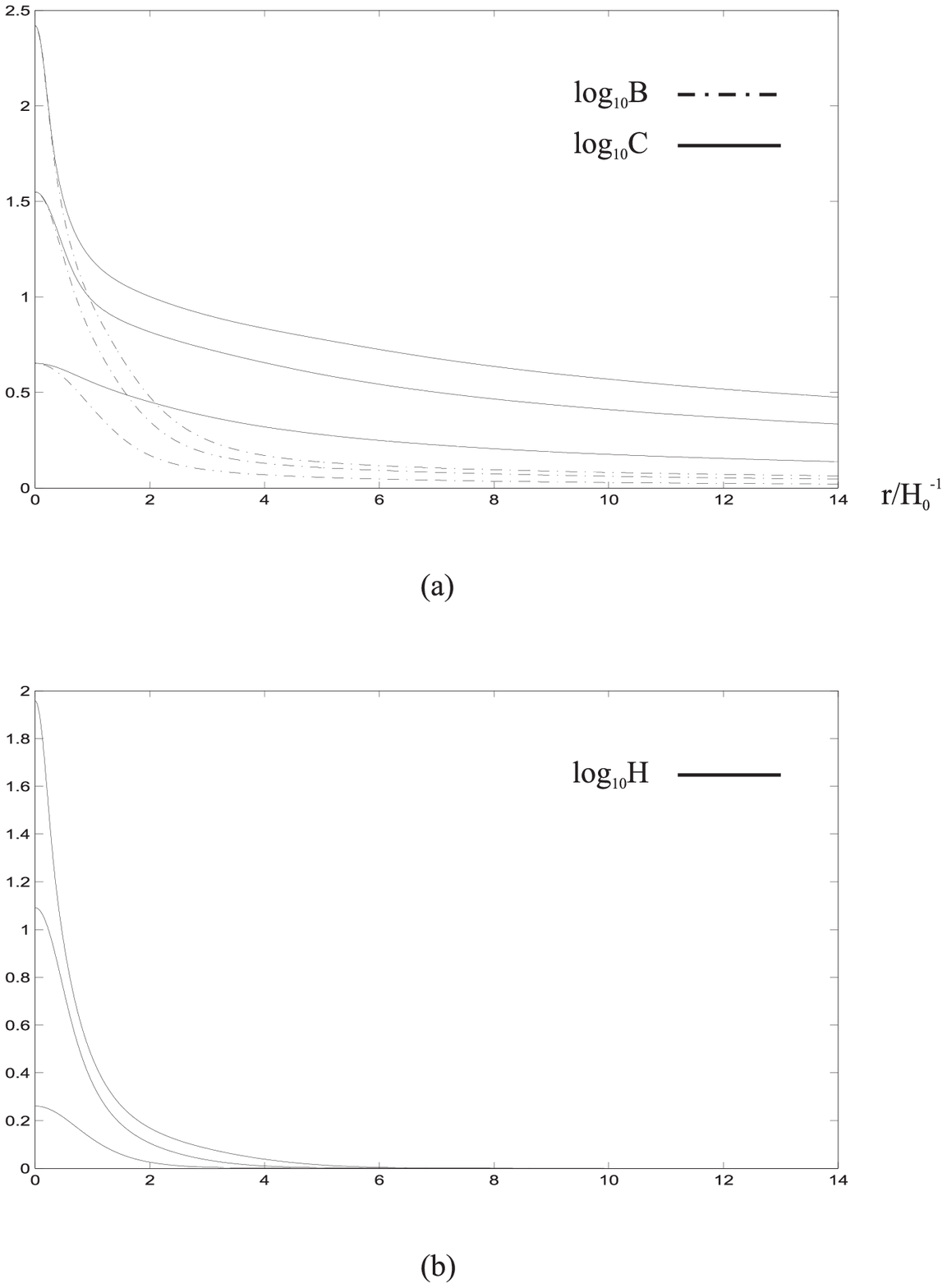}
\end{figure}

\newpage
\begin{figure}
\psfig{file=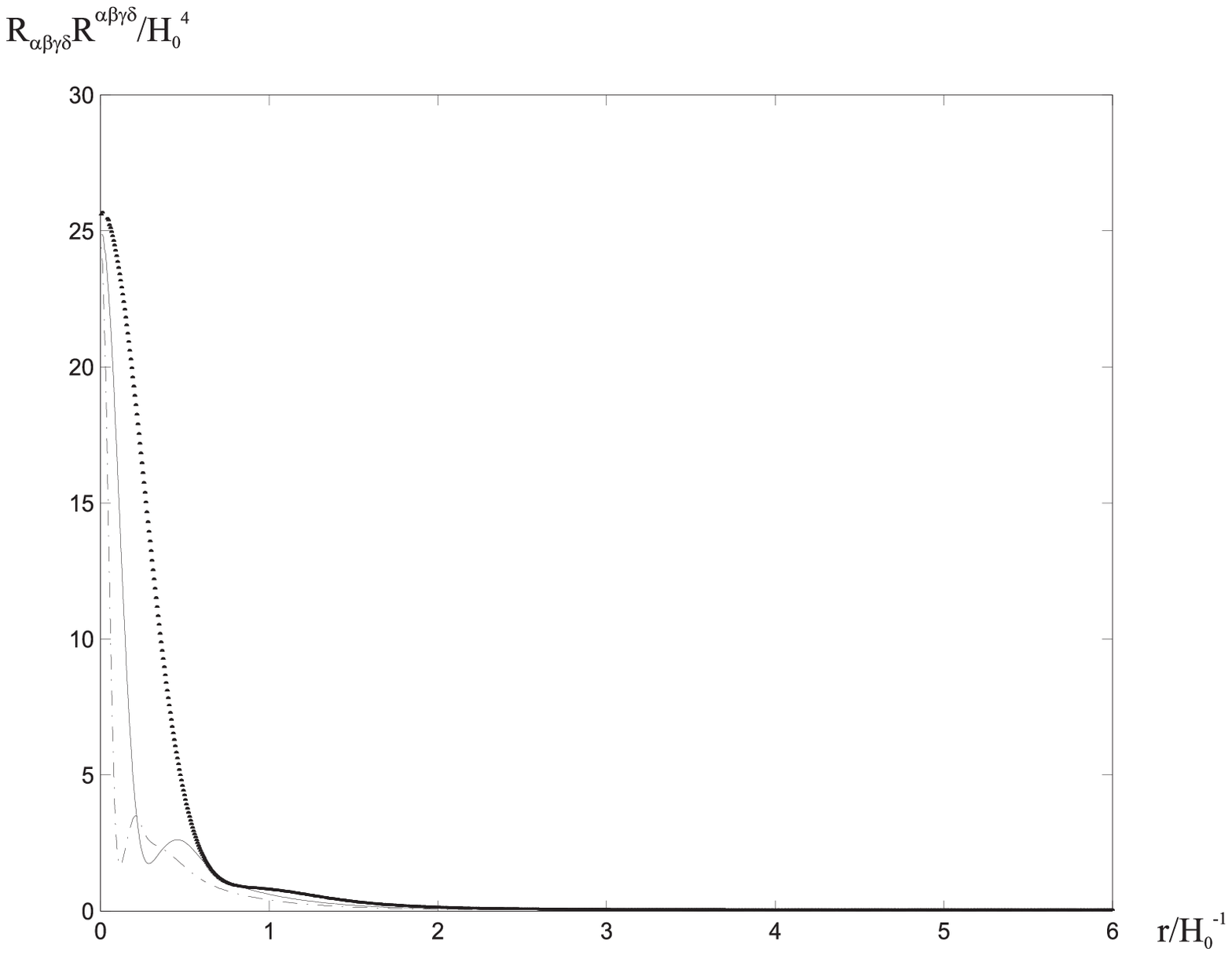}
\end{figure}

\end{document}